\documentstyle[twocolumn,prl,aps]{revtex}
\begin{document}
\draft
\title{ Ferromagnetic ground state of an orbital degenerate
electronic model for transition-metal oxides:
exact solution and physical mechanism}
\author{Shun-Qing Shen and Z. D. Wang}
\address{Department of Physics, The University of Hong Kong,
Pokfulam Road, Hong Kong, P. R. China}
\date{July 16, 1998}
\twocolumn[\hsize\textwidth\columnwidth\hsize\csname@twocolumnfalse\endcsname
\maketitle
\begin{abstract}
We present an exact ground state solution of a one-dimensional electronic
model for transition-metal oxides in the strong coupling limit.
The model contains doubly degenerated orbit for itinerant electrons and the
Hund coupling between the itinerant electrons and localized spins. 
The ground
state is proven to be a full ferromagnet for any density of electrons. Our
model provides a rigorous example for metallic ferromagnetism in narrow band
systems. 
The physical mechanism for ferromagnetism and
its relevance to high-dimensional systems, like R$_{1-x}$X$_x$MnO$_3$,
are discussed. Due to the orbital degeneracy of itinerant electrons, the superexchange
coupling can be ferromagnetic rather than antiferromagnetic in the one-band case.
\end{abstract}
                                                   
]

The origin of ferromagnetism in narrow band systems is a long standing
problem 
and has recently attracted renewed attention. Essentially
speaking, there are three basic routes to ferromagnetism:
1). it is formulated in the Hubbard model for an intermediate
short-range Coulomb interaction with a pronounced peak in the density
of state
near the band edge \cite{Gutzwiller63};
2). the Hund's rule coupling in the presence of
the orbital degeneracy of itinerant electrons leads to ferromagnetism
\cite{Slater36}; and
3). the motion of electrons in the localized spin background forces the spins
to align
parallelled, {\it i.e.}, the double exchange mechanism \cite{Zener51}.
Significant progresses have been made in our understanding
of ferromagnetism in last several years. Recent reviews on this subject
are seen in Refs.\cite{Tasaki98,Vollhardt97,Shen98a} and references therein.
On the other hand, ferromagnetism in R$_{1-x}$X$_x$MnO$_3$
and related ordering states have stimulated extensive interests due to the
 phenomenon of the colossal
magnetoresistance. The physical origins are still lack of full
understanding. Electronic models for this family of materials
contain all the three factors which favor to ferromagnetism, and are ideal
candidates to test all physical mechanisms for
ferromagnetism. Various techniques are applied to investigate magnetism in
these models
\cite{Kugel73}.
Due to the complexity of the systems quite few rigorous
solution or rigorous results are obtained to the models. Usually rigorous results are very helpful
to shape the physics of theoretical models, especially in such a strong correlated
electron system.

In this Letter we shall present an exact solution of the ground state for
a one-dimensional electronic model for transition-metal oxides in the case of strong
coupling. The ground states are proven to be ferromagnetic by means of the
Perron-Frobenius theorem. The physical mechanism
and its relevance to three-dimensional cases are discussed. Due to the
orbital degeneracy of itinerant electrons, the superexchange coupling can be
ferromagnetic rather than antiferromagnetic in the one-band case.

An electronic model Hamiltonian for the transition-metal oxides
is defined on a discrete lattice $\wedge$ with $N_{\wedge}$ site
and is
written as
\begin{equation}
H = H_t + H_u + H_j + H_h
\end{equation}
where
\begin{eqnarray}
H_t &=& \sum_{ij, \gamma,\gamma',\sigma}
t_{ij}^{\gamma\gamma'}
c^{\dagger}_{i,\gamma,\sigma} c_{j,\gamma',\sigma}; \nonumber \\
H_u &=& \sum_{i, \gamma,\gamma',\sigma,\sigma'}
(1-\delta_{\gamma,\gamma'}\delta_{\sigma,\sigma'})
U_{\gamma\gamma'}n_{i,\gamma,\sigma}n_{i,\gamma',\sigma'};\nonumber \\
H_j &=& -\sum_{i,\gamma\neq \gamma', \sigma,\sigma'} J_{\gamma\gamma'}
(c^{\dagger}_{i,\gamma,\sigma} c_{i,\gamma,\sigma'}
c^{\dagger}_{i,\gamma',\sigma'} c_{j,\gamma',\sigma}  \nonumber \\
& &+
c^{\dagger}_{i,\gamma,\sigma} c_{i,\gamma',\sigma}
c^{\dagger}_{i,\gamma,\sigma'} c_{i,\gamma',\sigma'});\nonumber \\
H_h &=& -\sum_{i,\gamma}J_H {\bf S}_i\cdot{\bf S}_{i,\gamma}.\nonumber
\end{eqnarray}
$c^{\dagger}_{i,\gamma,\sigma}$ and
$c_{i,\gamma,\sigma}$ are creation and annihilation
operators for $e_g$ electron with spin $\sigma$ ($=\pm 1$) on 
orbital $\gamma$($=\pm 1$) at site $i$, respectively.
${\bf S}_i$
is the maximal total spin
of three $t_{2g}$ electrons ($S=3/2$), and ${\bf S}_{i,\gamma} =
\sum_{\sigma\sigma'} \hat{\sigma}_{\sigma\sigma'} 
c^{\dagger}_{i,\gamma,\sigma}c_{i,\gamma,\sigma'}/2$ is a spin operator for 
an $e_g$ electron and $\hat{\sigma}$ are the Pauli matrices.
$H_t$ describes
the process of electrons hopping between nearest neighbor sites.
$H_u$ is of the
the on-site Coulomb interaction. $H_j$ is the Hund exchange coupling between
itinerant electrons on different orbital at the same site and
$J_{\gamma\gamma'}$
is always positive. $H_h$ is the Hund coupling between localized spin and
itinerant electrons at the same sites.
In this model the total spin operator,
$${\bf S}_{tot} = \sum_{i\in \wedge} {\bf S}_{i}
+ \sum_{i,\gamma} {\bf S}_{i,\gamma},$$
commutes with the Hamiltonian, $[{\bf S}_{tot}, H] =0$.
Hence the total spin and its z-component are good quantum numbers. The maximal
total spin is $S_{max} = N_{\wedge} S + N_e/2$ ($N_e$ is the number of electrons
and we just consider $N_e < 2 N_{\wedge}$).
We call the state with $S_{max}$ a ferromagnet.

This model has
been investigated by many authors \cite{Kugel73}. To the best of our
knowledge,
there has been no exact solution for this model until now. 
In order to establishes some rigorous results for this model we first
consider a one-dimensional open chain and focus our attention on
the case that
(i). $t^{\gamma\gamma'}_{ij} = -t \delta_{\gamma,\gamma'}\delta_{i,j\pm 1}$
and $t>0$;
(ii). $U_{\gamma\gamma'} = U= +\infty$ if $\gamma =\gamma'$;
(iii). $U_{\gamma\gamma'}=U'$ if $\gamma \neq \gamma'$; 
(iv). $J_{\gamma\gamma'}=J >0$ for $\gamma\neq \gamma'$;
(v). $J_H >0$.
It is worth stressing that the condition (ii) excludes the double
occupancy of electrons on the same orbital at the same site, which is essential
to solve the one-dimensional model exactly.

To solve the model rigorously, we first investigate the ground state
properties by means of the Perron-Frobenius theorem \cite{Simon93}. The ground
state is proven to be non-degenerate and possesses the maximal total spin. In
this case we can write down the ground state wavefunction explicitly by utilizing
the Bethe ansatz. The solution is valid for any density of electrons.
Perron-Frobenius Theorem (for a real symmetric matrix) states:

{\it Let $M=\{
m_{ij}\}$ be a real, symmetric, and square matrix. If the matrix $M$ satisfies
the conditions: 
(i). all its off-diagonal matrix elements are non-positive,
$m_{ij} \leq 0$; 
(ii). any $i$ and $j$ are connected by the matrix, {\it i.e.},
we can always find an integer $n$ such that $(M^n)_{ij} \neq 0$,
then the lowest eigenvalue is nondegenerate and all elements of the 
corresponding eigenvector can be taken to be positive.}

This theorem was successfully applied to prove the existence of ferromagnetism in the
quantum double exchange model \cite{Kubo82}, the one-band Hubbard model
\cite{Tasaki89} and the orbitally degenerate Hubbard model \cite{Shen98b}.
In order to apply this theorem to the present
model (Eq.(1)), we have to choose a suitable set of basis to write the
Hamiltonian in the form of a real, symmetric and square matrix which
obeys the two conditions. Consider the system contains
$N_e$ electrons. As the z-component of total spin, $S_{tot}^z$,
is a good quantum number, the Hilbert space can be decomposed by $S_{tot}^z$.
 Assume $N_1$ electrons on $\gamma=-1$
orbit with spin $\sigma_1 \leq \sigma_2\leq \dots \leq \sigma_{N_1}$ are
located at sites $x_1 < x_2 <\dots <x_{N_1}$, and $N_2$ electrons on $\gamma=1$
orbit with spin $\sigma_{N_1 +1}\leq \sigma_{N_1 +2}
\leq \dots \leq \sigma_{N_1+N_2}$ are located at sites
$x_{N_1+1} < x_{N_1 +2} <\dots<x_{N_e}$. $x_m = 1,\dots,N_{\wedge}$ and $m_i=0,
\dots, 2S$. As $N_1 -N_2$ is also a good
quantum number, thus we choose specific $S_{tot}^z$, $N_1$ and $N_2$. (We shall
show the lowest energy state is located at $N_1=N_2$ if $N_e$ is even,
or $N_1=N_2\pm 1$ if $N_e$ is odd.)
Denote $\alpha$ be one of the configurations
of electrons and localized spins,
\begin{eqnarray}
\vert \alpha \rangle &=&
c^{\dagger}_{x_1, -1, \sigma_1} \dots c^{\dagger}_{x_{N_1}, -1,\sigma_{N_1}}
c^{\dagger}_{x_{N_1+1}, 1, \sigma_{N_1+1}}
\dots c^{\dagger}_{x_{N_e}, 1,\sigma_{N_e}} \nonumber \\
&\times&({\bf S}_{1}^+)^{m_1} ({\bf S}_{2}^+)^{m_2}\dots
({\bf S}_{N_{\wedge}}^+)^{m_{N_{\wedge}}} \vert 0\rangle \nonumber
\end{eqnarray}
where the state $\vert 0\rangle $ has the properties:\\
1). $c_{x_m, \gamma_m,\sigma_m}\vert 0\rangle =0$ for any $m=1,\dots,N_e$;\\
2). ${\bf S}^-_{i}\vert 0\rangle =0$ for any $i=1,\dots,N_{\wedge}$.\\
As we just construct a basis within the subspace $S_{tot}^z$, it is required that
\begin{equation}
S_{tot}^z = \sum_{m=1}^{N_e}\frac{\sigma_m}{2} + \sum_{i=1}^{N_{\wedge}} (m_i -S).
\label{z-s}
\end{equation}
A complete set of basis consists of all possible configurations of
$\{ \{x_m\}, \{\gamma_m\}, \{\sigma_m\}, \{m_i\}\}$ with the condition
(\ref{z-s}). $\alpha$ in $\vert \alpha\rangle$ represents one of the possible
configurations.

On the basis we come to show that the Hamiltonian satisfies the
condition of non-positivity and connectivity.

(1). Non-positive off-diagonal elements: On the basis we choose, the Hamiltonian can be expressed
in the form of square matrix. All non-zero off-diagonal elements of the matrix
$\langle \alpha \vert H\vert \alpha'\rangle$ are:
(a) $-t$ if $x_n = x'_n\pm 1$ and all other indices are the same;
(b) $-J$ if $\sigma_n = \pm \sigma'_m$, $x_n=x'_m$, $\gamma_n = -\gamma'_m$ and
all other indices are the same;
(c) $-J_H$ if $m_i +\sigma_n = m'_i +\sigma'_m$,
$\gamma_n =\gamma'_m$, $x_n =x'_m =i$ and all other indices are the same. All
the non-zero elements are negative if $t, J, J_H > 0$. Therefore the Hamiltonian
matrix satisfies the first condition of the Perron-Frobenius theorem.

(2). Connectivity: all basis are connected through $H$: (a). the hopping
terms connect all lattice sites within the orbit $\gamma$; (b). the
Hund coupling $J$ connects the two orbits at the same site; (c). the Hund
coupling $J_H$ connects the orbits of itinerant electrons and the localized
spins. Combination of (a), (b) and (c) shows all basis are connected by
$H$. 

On the basis, the lowest energy state with $S_{tot}^z$ is expressed in the
form,
\begin{equation}
\vert\Psi\rangle =
\sum_{\alpha} f(\{x_m\}, \{\gamma_m\},\{\sigma_m\}, \{m_i\})
\vert \alpha\rangle.
\label{lowest}
\end{equation}
According to the Perron-Frobenius theorem, we conclude that
the lowest energy state with $S_{tot}^z$ is non-degenerate and
all coefficients $f$ can be chosen to be positive, $f>0$.

Since the lowest energy state with $S_{tot}^z$ (Eq.(\ref{lowest})) is
non-degenerate and the total spin is a good quantum number, therefore the state
must be an eigenstate of the total spin,
\begin{equation}
({\bf S}_{tot})^2 \vert \Psi\rangle = S_{tot}(S_{tot}+1)\vert \Psi\rangle.
\end{equation}
To determine $S_{tot}$, we first construct an eigenstate with the maximal total
spin $S_{max} = N_{\wedge} S + N_e/2$,
\begin{eqnarray}
& &\vert \Phi(S_{tot}^z=-N_{\wedge}S - N_e/2)\rangle = \nonumber \\
& &\sum_{\{x_m\}} c^{\dagger}_{x_1, -1, -1}
\dots c^{\dagger}_{x_{N_1}, -1,-1}
c^{\dagger}_{x_{N_1+1}, 1, -1}
\dots c^{\dagger}_{x_{N_e}, 1,-1}\vert 0\rangle.  \nonumber
\end{eqnarray}
The other $2S_{max}$ eigenstates with total spin $S_{max}$
and different z-components are expressed by utilizing spin SU(2)
symmetry,
\begin{eqnarray}
\vert\Phi(S_{tot}^z)\rangle &=&
 ({\bf S}_{tot}^+)^M \vert\Phi(S_{tot}^z=-N_{\wedge}S -N_e/2)\rangle \nonumber \\
 &=& \sum_{\alpha} 
 \vert \alpha\rangle
 \label{ferro}
 \end{eqnarray}
 where $M = S_{tot}^z + N_{\wedge}S +N_e/2$ and the summation runs over
 all possible configurations in the
 subspace of $S_{tot}^z$. The main feature of the state is
 that all coefficients on the basis in the subspace $S_{tot}^z$ are equal.
 This state is not orthogonal to
 the lowest energy state $\vert\Psi\rangle$ if they have the same z-component
 of total spin,
 \begin{equation}
 \langle \Psi\vert\Phi\rangle \neq 0,
 \end{equation}
since all coefficients in $\vert\Psi\rangle$ are positive. As both states
are eigenstates of the total spin ${\bf S}_{tot}$, we have
\begin{eqnarray}
\langle\Psi\vert ({\bf S}_{tot})^2\vert \Phi\rangle
&=& S_{tot}(S_{tot}+1)  \langle \Psi\vert\Phi\rangle \nonumber \\
&=& S_{max}(S_{max}+1) \langle \Psi\vert\Phi\rangle.
\end{eqnarray}
Hence we conclude that
$S_{tot}=S_{max},$
{\it i.e.}, the lowest energy state possesses the maximal total spin. In other word,
the state is fully ferromagnetic.

The lowest energy state with $S_{tot}^z$ is fully ferromagnetic. According
to spin SU(2) symmetry of the model, the ground state is $(2S_{max}+1)$-fold
degenerate. This property makes it possible to write down the ground state
wavefunction explicitly. The coefficients in the ground state are independent of
spin indices, {\it i.e.},
\begin{equation}
f(\{x_m\}, \{\gamma_m\},\{\sigma_m\}, \{m_i\})
=g(\{x_m\}, \{\gamma_m\}).
\end{equation}
Hence the Schr\"{o}dinger equation,
$H\vert\Psi\rangle =E_g\vert\Psi\rangle,$
is reduced to
\begin{eqnarray}
&-& t\sum_{m, \delta=\pm 1} g(\{x_1,\dots,x_m+\delta,\dots,x_{N_e}\},
\{\gamma_m\}) \nonumber \\
&+& U_{eff}\sum_{n<m} \delta_{x_n,x_m}(1-\delta_{\gamma_n,\gamma_m})
g(\{x_m\}, \{\gamma_m\})\nonumber \\
&=& (E_g -N_e J/2- N_{\wedge}s) g(\{x_m\}, \{\gamma_m\}).
\end{eqnarray}
where $E_g$ is ground state energy of the model. In the case of ferromagnetism,
the Schr\"{o}dinger equation is reduced to a one-band Hubbard model with
the on-site Coulomb interaction, $U_{eff} = U' -J$ if we use the orbit
indices instead of usual spin indices. This equation can be solved exactly by
means of the Bethe ansatz \cite{Lieb68}. The solution to $g$ is expressed as
\begin{equation}
g(\{x_m\}, \{\gamma_m\}) = \sum_{P} [Q,P] \exp[i\sum_{m=1}^{N_e} k_{P_m}
x_{Q_m}],
\label{lieb-wu}
\end{equation}
where $P$ and $Q$ are two permutations of $(1, 2, \dots, N_e)$. The coefficients
$[Q,P]$ have the relation,
\begin{eqnarray}
[Q,P] &=& Y^{i,i+1}_{nm}[Q,P'];\label{lieb-1}\\
Y^{i,i+1}_{nm} &=& \frac{(\sin k_n -\sin k_m)P^{i,i+1} - i U_{eff}/2}
{(\sin k_n -\sin k_m) + i U_{eff}/2} \label{lieb-2}
\end{eqnarray}
where
\begin{eqnarray}
P=(P_1, \dots P_i=n, P_{i+1}=m, \dots P_N);\nonumber \\
P'=(P_1, \dots P'_i=m, P'_{i+1}=n, \dots P_N).\nonumber
\end{eqnarray}
$k_n$ ($n=1,\dots,N_{\wedge}$) are determined by Eqs.(\ref{lieb-1})
and (\ref{lieb-2}). The ground state is located in the subspace
$N_1=N_2$ if the number of electrons is even,
or $N_1 = N_2 \pm 1$ if the number of electrons is odd.
Its lowest energy is
\begin{equation}
E_g = -2t\sum_{n=1}^{N_e}\cos k_n
-\frac{1}{2}N_e J - N_{\wedge} J_H S.
\end{equation}
The properties of the Lieb-Wu's solution have been discussed
extensively \cite{Korepin94}.
Except for the filling $N_e =N_{\wedge}$, there is no energy gap. Hence the
ground state is a metallic ferromagnet.

We have obtained the exact solution of the ground state of the electronic model
for transition metal oxides in a one-dimensional open chain
for any density of electrons in the case of $U=+\infty$.
The state is proven to be fully ferromagnetic. Here
we come to discuss the physical mechanism of ferromagnetism and its
possible relevance to realistic and three-dimensional systems.
The model we discuss contains all possible factors which favor
to ferromagnetism: large on-site Coulomb interaction, orbital degeneracy
of itinerant electrons and the Hund's rule coupling. In the case of
$J=J_H=0$, the model has highly spin degenerated ground state
when $U=+\infty$. Any finite $U$ will remove the degeneracy to form a
spin singlet state. Thus without the Hund's rule coupling the only
strong on-site
Coulomb interaction cannot drive electrons to form ferromagnetism. On the contrary,
the non-zero Hund's rule coupling, {\it i.e.}, $J\neq 0$ and $J_H\neq 0$, will
also remove the degeneracy to form a ferromagnet. This indicates that the interplay
between the on-site Coulomb interaction and the Hund's rule coupling plays an
essential role in the stability of ferromagnetism in the model.

The ferromagnetism may also be survived in the case of finite $U$ and
large $J$ and $J_H$. To see this effect, we consider a more realistic
and three-dimensional model which is used extensively to
describe the doped lanthanum manganese
oxides. The final consequences are also valid
to the one-dimensional case.  The transfer integrals in the model for
R$_{1-x}$X$_x$MnO$_3$ are assumed to take a Slater-Koster form given
by the hybridization between $e_g$ orbit and nearest oxygen $p$ orbit
\cite{Shiba97},
\begin{equation}
t_{ij}^{\gamma\gamma'}=-t \tau^x_{\gamma\gamma'},
-t \tau^y_{\gamma\gamma'}, -t\tau^z_{\gamma\gamma'}
\end{equation}
for ${\bf r}_j = {\bf r}_i\pm \hat{x}, {\bf r}_i\pm \hat{y}$
and ${\bf r}_i\pm \hat{z}$, respectively,
where
\[
\tau^x = \left ( \begin{array}{cc}
\frac{1}{4} & -\frac{\sqrt{3}}{4}\\
-\frac{\sqrt{3}}{4} & \frac{3}{4}
\end{array}
\right ),
\tau^y =
\left ( \begin{array}{cc}
\frac{1}{4} & \frac{\sqrt{3}}{4}\\
\frac{\sqrt{3}}{4} & \frac{3}{4}
\end{array}
 \right ),
 \tau^z =
\left ( \begin{array}{cc}
1& 0\\
0& 0
\end{array}
\right ).
\]
$\gamma=\pm 1$ represent
$\left ( \begin{array}{l} 1 \\ 0 \end{array} \right )= 3z^2 -r^2$
orbit and $\left ( \begin{array}{l} 0 \\ 1 \end{array} \right )=x^2-y^2$
orbit of $e_g$ electron, respectively.
In reality, ,
the parameters of the model
for the manganites are roughly estimated as $U\gg J_H, J \gg t$
\cite{Satpathy96}.
 Hence we
can obtain an effective Hamiltonian at $N_e=N_{\wedge}$
by means of combination of the projection technique and the 
perturbation technique \cite{Shen98c}
\begin{eqnarray}
&H&_{eff} =\nonumber \\
&-& \frac{t^2}{U'-\frac{J}{2}} \sum_{ij} \left (
\frac{{\cal S}_i\cdot {\cal S}_j +(S+1/2)(S+3/2)}{(S+1/2)(S+3/2)}\right )
P_{ij}^d
\nonumber \\
&+&\frac{t^2}{U'+\frac{3J}{2} +J_HS} \sum_{ij}\left (
\frac{{\cal S}_i\cdot {\cal S}_j -(S+1/2)^2}{(S+1/2)(S+3/2)}\right )
P_{ij}^d
\nonumber \\
&+& \frac{t^2}{U + J_HS} \sum_{ij} \left (
\frac{{\cal S}_i\cdot {\cal S}_j - (S+1/2)^2}{(S+1/2)^2}\right )
P_{ij}^s
\end{eqnarray}
where ${\cal S}_i$ is a spin operator with $S+1/2$ as a ferromagnetic
combination of the localized spin and itinerant electron at the same site.
$P_{ij}^{s(d)}$ is the projection operator for the orbital occupancy
\begin{eqnarray}
P_{ij}^s =\tau^{\alpha}_i\tau^{\alpha}_j;\nonumber \\
P_{ij}^d =\tau^{\alpha}_i(1-\tau^{\alpha}_j) \nonumber
\end{eqnarray}
with $\alpha= x,y,z$, which depends on the direction of ${\bf r}_i -{\bf r}_j$.
From the effective Hamiltonian, the first term favors to ferromagnetic
correlation and other two terms favor to antiferromagnetic correlation. When
$U=+\infty$ or sufficiently large, the third term is suppressed. Comparing the first
and second terms, we find that the ferromagnetic coupling is always stronger
than antiferromagnetic coupling since
\begin{equation}
\frac{t^2}{U'-\frac{J}{2}} > \frac{t^2}{U'+\frac{3J}{2} +J_HS}
\end{equation}
and $P^d_{ij}$ always has nonnegative eigenvalues. When $J_H S$ is sufficiently large and $U'-J/4$ is kept
unchanged, the first term is
predominant even for a finite $U$. However when $U$ decrease to finite, an antiferromagnetic
frustrations is introduced. The stability of ferromagnetism depends on the
competition between the strong Hund coupling and the effect of finite $U$.
In LaMnO$_3$, which $N_e=N_{\wedge}$, a
ferromagnetic and insulating phase was observed
at low temperatures \cite{Schiffer95}. This experimental observation is
in qualitatively agreement with our analysis. Oppositely, if we neglect the orbital
degeneracy of $e_g$ electrons and use a one-band Kondo lattice
model to describe the sample, we always have an antiferromagnetic phase
at low temperatures \cite{Shen96}. It implies that the orbital degree of freedom
is very important for us to understand the phase diagrams of doped manganese
oxides. When the system is doped, {\it i.e.} $N_e < N_{\wedge}$, due to the strong
Hund coupling, $J_H \gg t$, the motion of itinerant electrons tends to force
the localized spins to align parallelled. Thus the double exchange mechanism as well
as the orbital degeneracy and strong $U$ should be also responsible for the metallic
ferromagnetism in R$_{1-x}$X$_x$MnO$_3$. The detailed discussion of the phase
diagrams will be published elsewhere.

Summarizing, the model on an open chain is solvable in the case of
$U=+\infty$. The ground state is metallic ferromagnetic except
for the density $N_e/N_{\wedge} =1$. The charge and orbital degrees of
freedom
are determined by a Lieb-Wu solution for one-dimensional Hubbard model.
The ferromagnetism can be survived in finite $U$ and higher dimensional
system. Due to the orbital degeneracy of itinerant
electrons and strong on-site repulsion $U$, the superexchange coupling can be ferromagnetic rather than
antiferromagnetic in the one-band case.

This work was supported by a CRCG research grant at the University of Hong
Kong.


\begin{references}
\bibitem{Gutzwiller63}
M. C. Gutzwiller, Phys. Rev. Lett. {\bf 10}, 59 (1963);
J. Hubbard, Proc. Roy. Soc. London A {\bf 276}, 238 (1963);
J. Kanamori, Prog. Theor. Phys. {\bf 30}, 275 (1963).
\bibitem{Slater36}
J. C. Slater, Phys. Rev. {\bf 49}, 537 (1936);
J. H. van Vleck, Rev. Mod. Phys. {\bf 25}, 220 (1953);
L. M. Roth, Phys. Rev. {\bf 149}, 306 (1966).
\bibitem{Zener51}
C. Zener, Phys. Rev. {\bf 82}, 403 (1951); P. W. Anderson and H. Hasegawa,
Phys. Rev. {\bf 100}, 675 (1955),
K. Kubo and N. Ohata, J. Phys. Soc. Jpn. {\bf 33}, 21 (1972).
\bibitem{Tasaki98}                                       
H. Tasaki, Prog. Theor. Phys. {\bf 99}, 489 (1998).
\bibitem{Vollhardt97}
D. Vollhardt, N. B\"{u}mer, K. Held, M. Kollar, J. Schlipf, and M. Ulmke,
Z. Phys. B {\bf 103}, 283 (1997).
\bibitem{Shen98a}
S. Q. Shen, Int. J. Mod. Phys. B {\bf 12}, 709 (1998).
\bibitem{Kugel73}      
K. I. Kugel and D. I. Khomskii, Sov. Phys. JETP {\bf 37}, 725 (1973);
S. Ishihara, J. Inoue, and S. Maekawa, Physica C {\bf 263}, 130 (1996);
Phys. Rev. B {\bf 55}, 8280 (1997);
R. Shiina, T. Nishitani, and H. Shiba, 
J. Phys. Soc. Jpn {\bf 66}, 3159 (1997);
S. Ishihara, M. Yamanaka, and N. Nagaosa, Phys. Rev. B {\bf 56}, 686 (1997).
\bibitem{Simon93}                                
See B. Simon, Statistical mechanics of lattice gas (Princeton University Press,
Princeton, New Jersey, 1993).
\bibitem{Kubo82}
K. Kubo, J. Phys. Soc. Jpn {\bf 51}, 782 (1982).
\bibitem{Tasaki89}
H. Tasaki, Phys. Rev. B {\bf 40}, 9192 (1989).
\bibitem{Shen98b}
S. Q. Shen, Phys. Rev. B {\bf 57}, 6474 (1998).
\bibitem{Lieb68}
E. H. Lieb and F. Y. Wu, Phys. Rev. Lett. {\bf 20}, 1445 (1968).
\bibitem{Korepin94}
See V. E. Korepin and F. H. L. Essler, Exactly Solvable Models of
 Strongly Correlated Electrons, (World Scientific, Singapore, 1994). 
\bibitem{Shiba97}
H. Shiba, R. Shiina, and A. Takahashi,
J. Phys. Soc. Jpn. {\bf 66}, 941 (1997).
\bibitem{Satpathy96}                   
S. Satpathy, Z. S. Popovi\'{c}, and F. R. Vukajlovi\'{c},
Phys. Rev. Lett.{\bf 76}, 960 (1996);
S. K. Mishra, R. Pandit, and S. Satpathy,
Phys. Rev. B {\bf 56}, 3316 (1996).
\bibitem{Shen98c}
S. Q. Shen and Z. D. Wang, unpublished.
\bibitem{Schiffer95}
J. B. Goodenough, Phys. Rev. {\bf 100}, 564 (1955);
P. Schiffer {\it et  al.}, Phys. Rev. Lett. {\bf 75}, 3336 (1995).
\bibitem{Shen96}
S. Q. Shen, Phys. Rev. B {\bf 53}, 14252 (1996); {\bf 55}, 14330 (1997).
\end{references}
\end{document}